\documentclass[journal]{IEEEtran}
\usepackage{amsmath}
\usepackage{epsfig}
\usepackage{graphicx}
\usepackage{subfigure}
\usepackage{multirow}
\usepackage{color}
\usepackage{changepage}
\usepackage{times}
\usepackage{float}
\usepackage{amssymb}
\usepackage{bbding}
\usepackage{booktabs}
\usepackage{caption}
\usepackage{cite}
\usepackage{bm}
\usepackage{array}
\usepackage{tabu}
\usepackage{soul}
\usepackage{url}
\usepackage{tcolorbox}

\soulregister\cite7
\soulregister\ref7 

\ifCLASSINFOpdf
\else
\fi
\begin{document}
%
\title{Angel's Girl for Blind Painters: an Efficient Painting Navigation System Validated by Multimodal Evaluation Approach}
%
%
\author{Hang Liu,
        Menghan Hu,
        Yuzhen Chen,
        Qingli Li,~\IEEEmembership{Senior Member,~IEEE,}
        Guangtao Zhai,~\IEEEmembership{Senior Member,~IEEE,}
        Simon X. Yang,~\IEEEmembership{Senior Member,~IEEE,}
        Xiao-Ping Zhang,~\IEEEmembership{Fellow,~IEEE,}
        Xiaokang Yang,~\IEEEmembership{Fellow,~IEEE}
\thanks{This work is sponsored by the Shanghai Sailing Program (No. 19YF1414100), the National Natural Science Foundation of China (No. 61831015, No. 61901172), the STCSM (No. 18DZ2270700), the Science and Technology Commission of Shanghai Municipality (No. 18511102500), and “Chenguang Program” supported by Shanghai Education Development Foundation and Shanghai Municipal Education Commission (No. 19CG27).}
\thanks{H. Liu, M. Hu, Y. Chen and Q. Li are with the Shanghai Key Laboratory of Multidimensional Information Processing, East China Normal University.}
\thanks{M. Hu, G. Zhai, and X. Yang are with the Key Laboratory of Articial Intelligence, Ministry of Education.}
\thanks{Simon X. Yang is with the Advanced Robotics and Intelligent Systems Laboratory, School of Engineering, University of Guelph.}
\thanks{Xiao-Ping Zhang is with the Department of Electrical, Computer and Biomedical Engineering, Ryerson University.}
\thanks{Corresponding author: Menghan Hu (mhhu@ce.ecnu.edu.cn)}}

%
%

\markboth{Journal of \LaTeX\ Class Files,~Vol.~1, No.~1, August~2021}%
{Shell \MakeLowercase{\textit{et al.}}: Bare Demo of IEEEtran.cls for IEEE Journals}
%



\maketitle

\begin{abstract}
 For people who ardently love painting but unfortunately have visual impairments, holding a paintbrush to create a work is a very difficult task. People in this special group are eager to pick up the paintbrush, like Leonardo da Vinci, to create and make full use of their own talents. Therefore, to maximally bridge this gap, we propose a painting navigation system to assist blind people in painting and artistic creation. The proposed system is composed of cognitive system and guidance system. The system adopts drawing board positioning based on QR code, brush navigation based on target detection and bush real-time positioning.
 Meanwhile, this paper uses human-computer interaction on the basis of voice and a simple but efficient position information coding rule. In addition, we design a criterion to efficiently judge whether the brush reaches the target or not. According to the experimental results, the thermal curves extracted from the faces of testers show that it is relatively well accepted by blindfolded and even blind testers. With the prompt frequency of 1s, the painting navigation system performs best with the completion degree of $89\%\pm8.37\%$ and overflow degree of $347\%\pm162.14\%$. Meanwhile, the excellent and good types of brush tip trajectory account for $74\%$, and the relative movement distance is $4.21\pm2.51$. This work demonstrates that it is practicable for the blind people to feel the world through the brush in their hands. In the future, we plan to deploy ``Angle's Eyes'' on the phone to make it more portable. The demo video of the proposed painting navigation system is available at:
\url{https://doi.org/10.6084/m9.figshare.9760004.v1}.
\end{abstract}


%
\IEEEpeerreviewmaketitle

\section{Introduction}
\label{sec:intro}
%
%
%
%

\IEEEPARstart{A}{ccording} to the recent report of the world health organization, about 1.3 billion people worldwide suffer from various forms of visual impairment, and about 36 million people are blind \cite{WHO,z}. Visual organ is one of the most important sensory organs of human beings. More than 90$\%$ of sensory information transmitted to human beings is visual \cite{zhang2019arcore,horton2017review}. For blind people, the decline or lack of visual perception brings great inconvenience to their daily life. They can hardly complete some routine tasks just like walking, driving, writing and recognizing objects \cite{Hu2019AN,7422089}. Vision loss can also greatly affect artistic creations such as painting. Though having dexterous hands, many of them find their artistic dreams hard to realize because of visual inconvenience. They are eager to pick up the paintbrush, like Leonardo da Vinci, to create and make full use of their own talents. Obstacles that mainly come from their visual impairments keep their dreams so far away from them!

Painting can not only mould a person’s temperament, but also improve communication between people \cite{lyon1995drawing}. Drawing images can assist the process of remembering information, which turns out to be a good memory strategy to quickly grasp a new concept \cite{wammes2016drawing,fernandes2018surprisingly} and help acquire knowledge of different subjects \cite{fan2015drawing,ainsworth2011drawing}. While painting, people can release their pent-up emotions as a way to maintain an optimistic attitude \cite{de2005does,drake2011short}. After having taken training in art, blind people can also better express themselves and adjust their mood during the process of painting. More benefits can be further obtained by blind people through artistic creation, such as the development of their brain areas related to visual memory \cite{cacciamani2017memory,likova2012drawing}.

White cane, dog guides and wearable vision assistance \cite{7122303, 8370108}, etc., have been used to help visually impaired and blind people  finish daily tasks. Recent years have seen a rapid technological development in the navigation system based on spatial orientation for blind people. The existing navigation systems to which technologies of route planning and human-computer interaction are usually applied can provide navigational aids for blind users  and protect them from a fall or a tumble \cite{jeamwatthanachai2019indoor, zhao2020learning}. Zhang et al. designed an indoor navigation system for blind people by using technologies of ARCore-based visual positioning and touch-based human-computer interaction \cite{zhang2019arcore}. Murata et al. designed a navigation assistance system for blind people using mobile phone positioning technology, which offers guidance when blind people walk around in a large-scale indoor public place \cite{murata2019smartphone}. Li et al. designed a new navigation system based on overall vision, which can help blind people themselves move about while being indoors \cite{li2018vision}. Cordeiro et al. developed a high level information fusion technology in their navigation system and a system can automatically calculate the possibility of collision in the moving process of blind users \cite{cordeiro2019collision}.

Most of the current researches focus on how to use robot sense technologies (computer vision, ultrasonic sensing, etc.) to help blind people safely and successfully perform basic daily tasks such as positioning \cite{murata2019smartphone,zhang2019arcore,pravin2018vlc} and obstacle avoidance \cite{cordeiro2019collision,martinez2017using,kolarik2017blindness}. As far as we know, there are few studies that employ sensing technologies such as computer vision to assist blind people in accomplishing higher-level tasks, including painting.

Traditional drawing auxiliaries for blind people use their tactile perception and low-tech analog tools to draw images \cite{bornschein2017digital,prescher2014production}. Vinter et al. designed a drawing board made out of thin plastic, which is based on tactile perception \cite{vinter2012exploratory}. With a pen, the board enables blind people to draw images, though it can only help create simple geometric patterns due to the low tactile perception dimension (it is one-dimensional to our knowledge). Thus it is difficult to help create more vivid images. Kamel and Landay designed a simple static grid to help blind users, while painting, determine the paper’s locating points from the start point to the end point \cite{kamel1999integrated}. The positioning scheme turns the paper into the uneven surface paper, which may eventually have a negative impact on the painting. On the other hand, it is relatively complicated for blind people to operate owing to the absence of feedback. Concomitant with the development of science and technology, researchers developed drawing equipment for blind people based on tactile display. Recently, Bornschein et al. have designed an array display for blind people to make tactile drawings by means of two-dimensional tactile input and output matrices \cite{bornschein2018comparing}. Tactile display devices can interactively guide blind people in drawing, but they are expensive and difficult to operate. With protruding points on the drawing paper affecting user experience, the navigation system for blind people, which takes tactile sense as the core solution, can only assist blind users to draw simple and low-dimensional images. In view of the defects of tactile navigation system for painting, we intend to design a navigation system for painting using visual assistance, which may enhance blind people’s dimension of perception with regard to painting space, so as to help add more artistic appeal to the painting and bring more pleasures of creation.

Therefore, the main contribution of the current work is that we develop an efficient painting navigation system based on visual aid and auditory feedback for blind people to help them with painting training and artistic creation. This painting navigation system is able to serve as an ``Angle's Eyes'' of blind painters, and provide the possibility for them to feel the world through the brush in their hands.

Aiming at overcoming the difficulties of the painting navigation system for blind people, we have carried out the following work: (1) to realize the real-time monitoring and correction of the painting position, we adopt the QR code positioning technology; (2) to accurately and efficiently identify brush tip position, we utilize the dynamic brush tip recognition algorithm; (3) to locate the tip position in an efficient, accurate, and real-time manner, we divide the painting area using virtual grids. These virtual grids are simply coded, allowing blind users to easily access the tip location; (4) to achieve the human-machine information exchange when fulfilling drawing tasks, we develop a digestible two-way voice interaction method. Blind people can tell the navigation system where to paint via the pre-designed voice, and then follow the voice cue to find the targeted region; and (5) to efficiently test the performance of the painting navigation system, we use the multimodal evaluation metrics including recognition accuracy and user experience evaluation. In addition, to explore the navigation effect with different frequencies of voice prompt, three prompt frequencies namely fast, medium and slow are set during the comparative experiments.

\section{Design of painting navigation system}
\begin{figure}
  \centering
  \includegraphics[width=0.5 \textwidth]{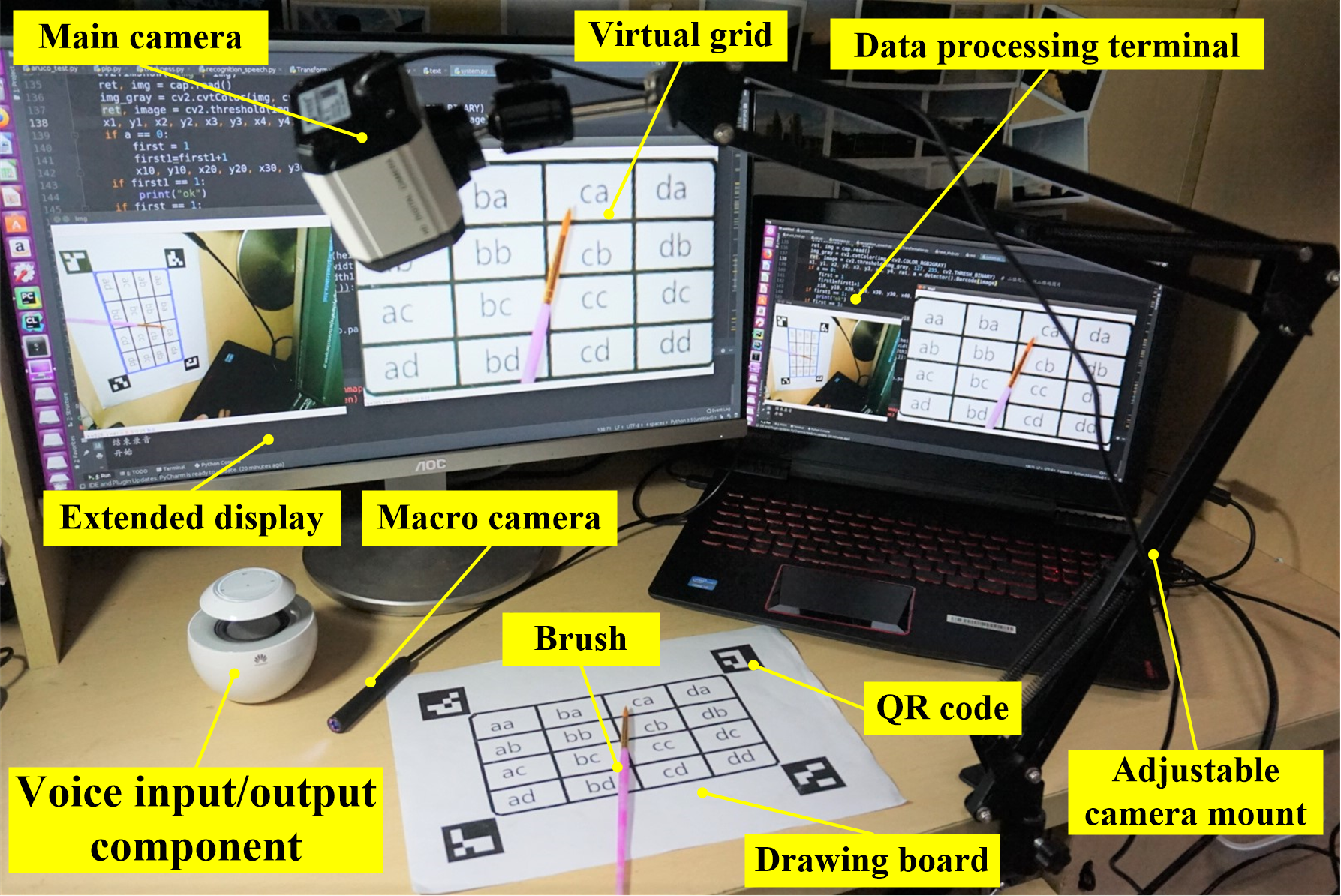}
  \caption{Setup of the painting navigation system for blind painters.}\label{A0}
\end{figure}

The photograph of painting navigation system is illustrated in Fig. \ref{A0}. The system mainly consists of drawing board, voice input/output component, main camera, macro camera, extended display, QR code, data processing terminal, adjustable camera mount, virtual grid and brush. The above components further constitute the other modules in the system, thereby realizing the functions of the painting navigation system. The demo video of the proposed painting navigation system is available at: \url{https://doi.org/10.6084/m9.figshare.9760004.v1}. In this demo video, the brush color recognition module is demonstrated. Considering the length of the paper and the high accuracy rate of the brush color recognition, this module is not described in this paper.

\subsection{Information Stream in Painting Navigation System}
\begin{figure}
\centering
  \includegraphics[width=0.4 \textwidth]{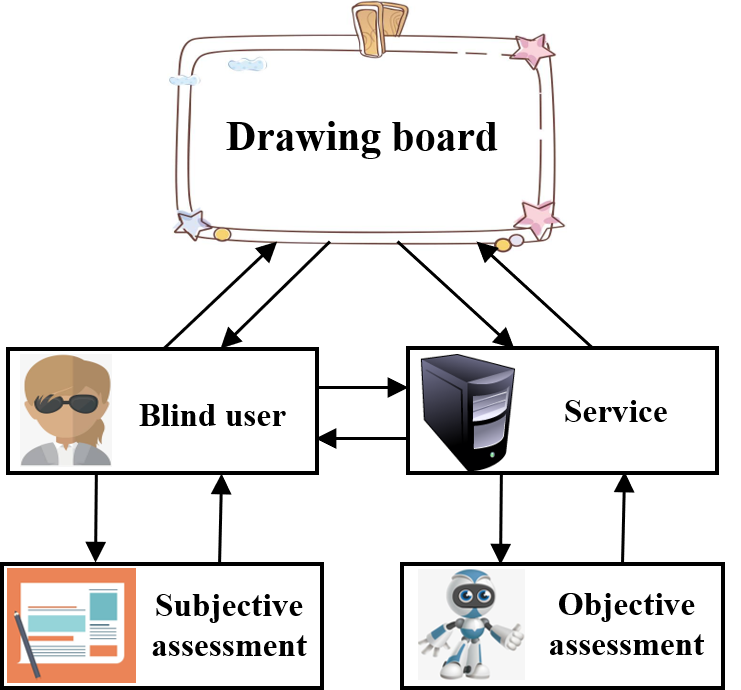}
  \caption{Main information flow in painting navigation system. The subjective assessment methods are used for evaluating the system performance on the user's feeling, and the objective assessment methods are used for evaluating the performance of each technique.}\label{A1}
\end{figure}

\begin{figure*}
  \centering
  \includegraphics[width=1 \textwidth]{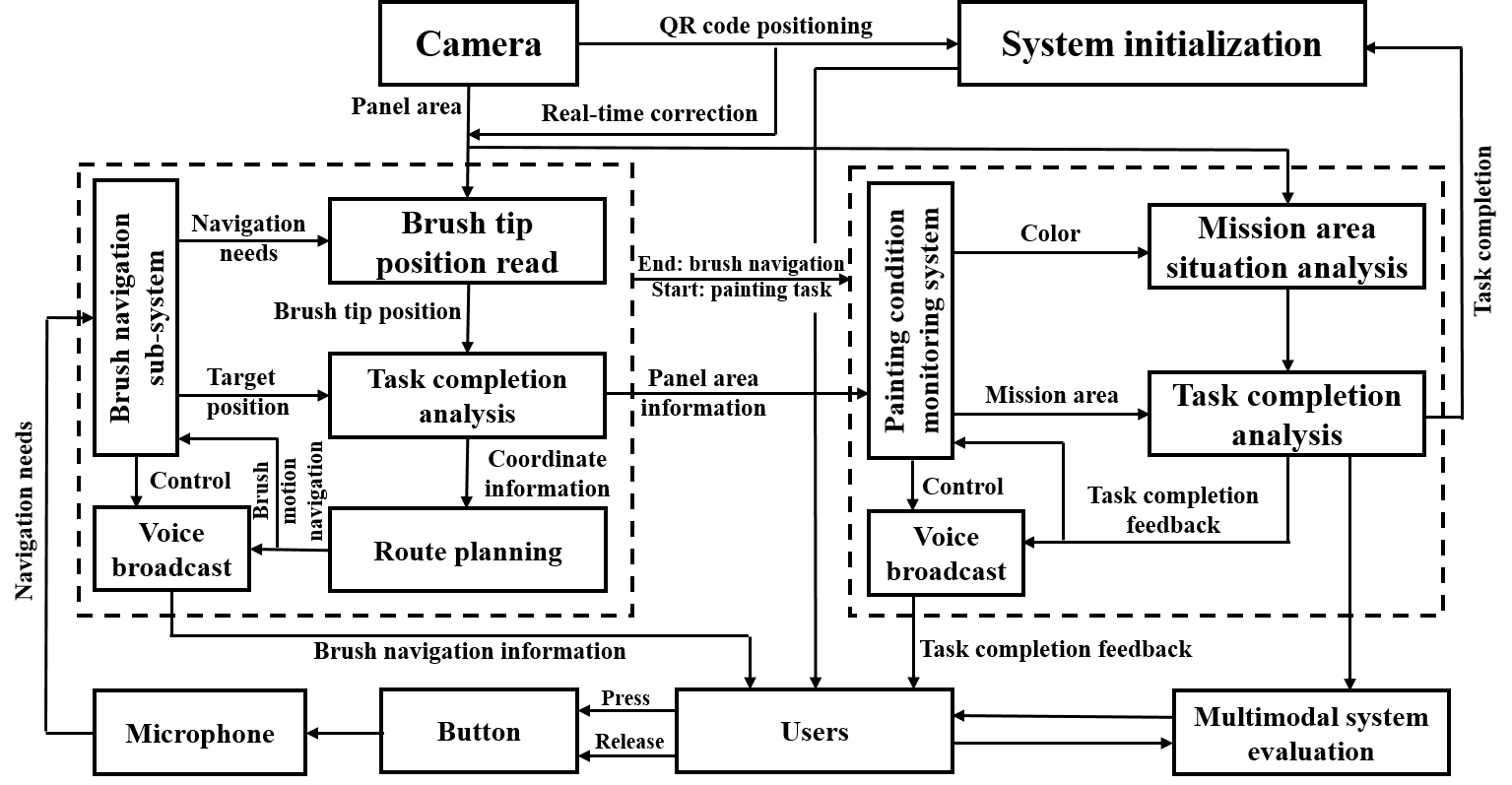}
  \caption{Detailed working logic among the painting navigation system.}\label{A2}
\end{figure*}

To complete the painting task, the painting navigation system needs to process multimodal and multilevel information. Consequently, in the process of system design, it is necessary to understand the information flow. Fig. \ref{A1} shows the path of the main information flow in the proposed painting navigation system. The main part of painting navigation system is cognitive system and guidance system. The cognitive system aims to help blind people understand the world near the drawing board (macro control of drawing area, dynamic recognition of brush position, access to the targeted drawing region, etc.), and it is composed of QR code based drawing board positioning module and brush real-time positioning module. The guidance system aims to build an informational bridge between the real world and users or services, and its main part is human-computer interaction module. As shown in Fig. \ref{A1}, we design the main working logic among the painting navigation system: locate the stable top view for the painting navigation system $\rightarrow$ recognise the user's voice command information $\rightarrow$ identify the coordinate information via semantic analysis $\rightarrow$ detect the brush position in real time $\rightarrow$ transmit location information of brush to users in real time. More detailed working logic diagram is shown in Fig. \ref{A2}.

\subsection{QR Code based Drawing Board Positioning Module}
\begin{figure*}
  \centering
  \includegraphics[width=1 \textwidth]{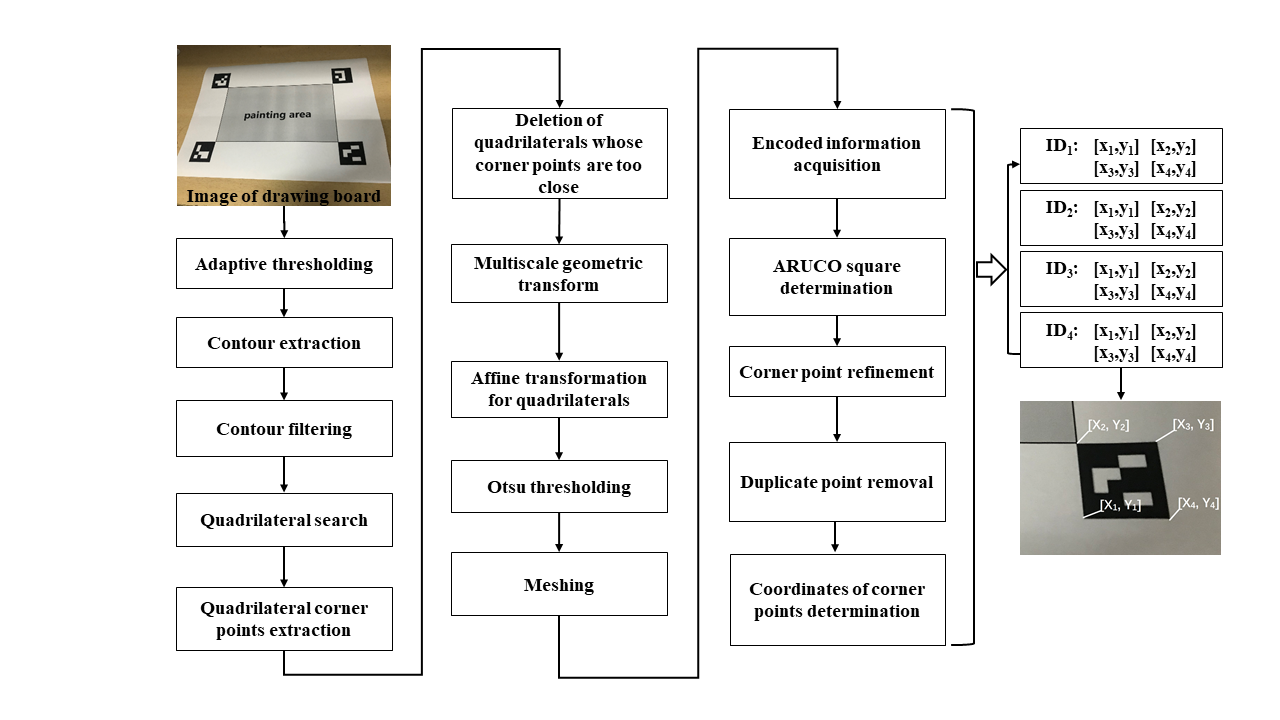}
  \caption{Procedure for identifying four QR codes and calculating their coordinates.}\label{A3}
\end{figure*}

To help the blind complete the painting task, the drawing board needs to be positioned. We combine the QR code positioning and the perspective transformation to achieve the positioning of the drawing board. Four ARUCO squares with different encoded information are printed on the four sides of the drawing board. ARUCO markers are synthetic square markers. ARUCO library was originally developed by Rafael Muñoz and Sergio Garrido \cite{Aruco1}. ARUCO markers comprise a wide black border with an inner binary matrix, so that they can be recognized quickly \cite{Aruco2}. The ARUCO squares are marked as [$ARUCO_{j}X_{1}$, $ARUCO_{j}Y_1$], [$ARUCO_{j}X_2$, $ARUCO_{j}Y_2$], [$ARUCO_{j}X_3$, $ARUCO_{j}Y_3$], and [$ARUCO_{j}X_4$, $ARUCO_{j}Y_4$], respectively, where $j$ $\in$ $\{$1,2,3,4$\}$. This is the encoding information of the ARUCO squares. After the system detects an ARUCO square with corresponding coded information, it will automatically recognize the four vertex coordinates of the ARUCO square. The designed drawing board is then imaged by the main camera. Assisted by QR code and perspective transformation algorithm, the drawing board can be located with any camera angles. The whole procedure for identifying four QR codes and calculating their coordinates is presented in Fig. \ref{A3}. According to the layout of the ARUCO squares, the corner of ARUCO square closest to the center is one of the final four location points, and therefore, the final coordinates of drawing board can be calculated using:

\begin{equation}\label{E00}
  [X_{i}, Y_{i}] = [ARUCO_{j}X_{k}, ARUCO_{j}Y_{k}]
\end{equation}
when $i$ value is 1, 2, 3, and 4, the corresponding $k$ value is 4, 3, 2, and 1, which respectively correspond to the upper left corner, upper right corner, left bottom corner, and right bottom corner of drawing board.

Subsequently, the painting region is registered using the perspective transformation. After this operation, we can realize the transformation of painting area from squint angle to overhead angle. Let assume that the coordinates of detected ARUCO squares in the original image and in the transformed image are respectively $O$=($x_{i}$,$y_{i}$) and $T$=($x_{i}^{'}$,$y_{i}^{'}$), where $i$ $\in$ $\{$1,2,3,4$\}$. The correlation between these two coordinates is $T$=$O$$\times$$M$, where $M$ is the transfer matrix.

The conversion formula of perspective transformation is:
\begin{equation}\label{E0001}
    [x^{'}\ y^{'}\ w^{'}] = [u\ v\ w]M = [u\ v\ w]\begin{bmatrix} a_{11} & a_{12}& a_{13} \\ a_{21} & a_{22}& a_{23}\\ a_{31} & a_{32}& a_{33} \end{bmatrix}
\end{equation}
where ($u$,$v$) is the coordinates from original image, ($x$, $y$) is the coordinates after conversion. According to formula 2, ($x$, $y$) can be expressed as:

\begin{footnotesize}
\begin{equation}\label{E0002}
    (x,y)=( \frac{x^{'}}{w^{'}}, \frac{y^{'}}{w^{'}})=( \frac{a_{11}u+a_{21}v+a_{31}}{a_{13}u+a_{23}v+a_{33}},\frac{a_{12}u+a_{22}v+a_{32}}{a_{13}u+a_{23}v+a_{33}})
\end{equation}
\end{footnotesize}

Let $w = 1$, $a_{31}=1$,$a_{32}=1$, $a_{33}=1$, $M$ can be expressed as:

\begin{equation}\label{E0003}
    M=\begin{bmatrix} a_{11} & a_{12}& a_{13} \\ a_{21} & a_{22}& a_{23}\\ a_{31} & a_{32}& a_{33} \end{bmatrix}
\end{equation}

Therefore, we can calculate $M_{1}$ in transformation ($x_{0}$, $y_{0}$)$\rightarrow${}(0, 0), ($x_{1}$, $y_{1}$)$\rightarrow${}(1, 0), ($x_{2}$, $y_{2}$)$\rightarrow${}(1, 1), ($x_{3}$, $y_{3}$)$\rightarrow${}(0, 1), and can calculate $M_{2}$ in transformation (0, 0)$\rightarrow${}($x_{0}^{’}$, $y_{0}^{’}$), (1, 0)$\rightarrow${}($x_{1}^{’}$, $y_{1}^{’}$), (1, 1)$\rightarrow${}($x_{2}^{’}$, $y_{2}^{’}$), (0, 1)$\rightarrow${}($x_{3}^{’}$, $y_{3}^{’}$). The transformation ($x_{0}$, $y_{0}$)$\rightarrow$($x_{0}^{’}$, $y_{0}^{’}$), ($x_{1}$, $y_{1}$)$\rightarrow$($x_{1}^{’}$, $y_{1}^{’}$), ($x_{2}$, $y_{2}$)$\rightarrow$($x_{2}^{’}$, $y_{2}^{’}$), ($x_{3}$, $y_{3}$)$\rightarrow$($x_{3}^{’}$, $y_{3}^{’}$) can be seen as: ($x_{0}$, $y_{0}$)$\rightarrow$(0, 0)$\rightarrow$($x_{0}^{’}$, $y_{0}^{’}$), ($x_{1}$, $y_{1}$) $\rightarrow$(1, 0)$\rightarrow$($x_{1}^{’}$, $y_{1}^{’}$), ($x_{2}$, $y_{2}$)$\rightarrow$(1, 1)$\rightarrow$($x_{2}^{’}$, $y_{2}^{’}$), ($x_{3}$, $y_{3}$)$\rightarrow$(0, 1)$\rightarrow$($x_{3}^{’}$, $y_{3}^{’}$). So that we can calculate $M$: $M$ = $M_{1}$$M_{2}$.


The coordinates of corner points determined by the image processing method are discrete observations, and they inevitably will change over time by the user intervention (move the camera or the panel). To reduce this error, when the camera detects the ARUCO squares simultaneously, the position of drawing board is updated.

\subsection{Brush Real-time Positioning Module}
\begin{figure*}
  \centering
  \includegraphics[width=1 \textwidth]{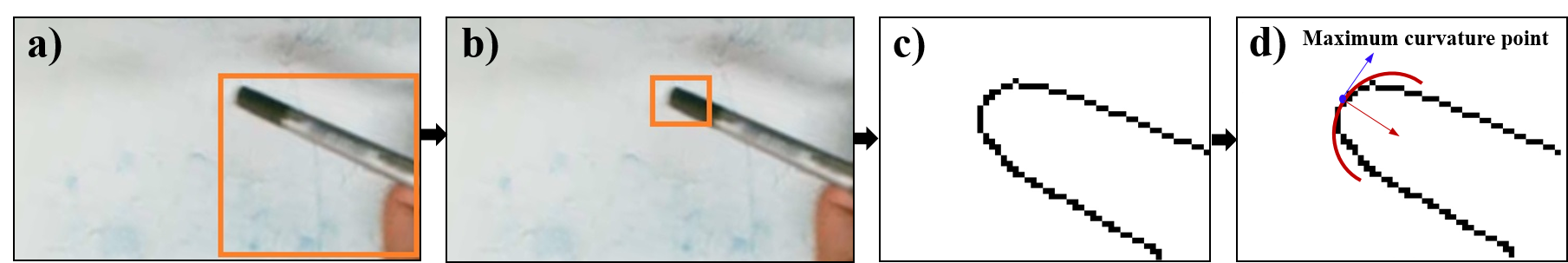}
  \caption{Process of brush tip detection: a) brush region identification using  dynamic target detection; b) brush tip detection; c) tip’s edge detection via Canny edge operator; and d) determination of tip’s center based on maximum curvature.}\label{A4}
\end{figure*}

To navigate the painting process, it is necessary to identify and locate the position of brush tip. The process is divided into four steps (Fig. \ref{A4}): 1) brush identification: the dynamic target recognition algorithm \cite{kaewtrakulpong2002improved} is adopted to identify the dynamic objects in the drawing area. In the process of painting, the drawing area is still, while the brush is dynamic. Therefore, the brush can be identified through the dynamic target recognition algorithm; 2) brush tip detection: YOLOv3 \cite{redmon2018yolov3} is used to accurately position the brush tip; 3) tip’s edge detection: Canny edge operator is applied to detect the nip's edge; and 4) tip’s center determination: the points on the edge are numbered in order. Each edge is a line made up of many points. We assume the total number of points that make up the line is $M$, and the serial number of the points is $i$. The curvature of $i$th point is presented as $Cur_i$, and it can be calculated using:

\begin{equation}\label{E1}
  Cur_i = \sqrt{(\sum_{k=-j}^{k=j}(x_{i+k}-x_{i}))^{2}+(\sum_{k=-j}^{k=j}(y_{i+k}-y_{i}))^{2}}
\end{equation}
where $j$ denotes the curvature calculation range, and $k=1,2,...,j$ is the serial number within the range determined by $j$. The value of $j$ can be assigned based on the scale of application scenario. That is, it mainly depends on the performance of the used camera and the distance between the used camera and panel. In this case, we set the $j$ as 5.

\begin{figure}
  \centering
  \includegraphics[width=0.45 \textwidth]{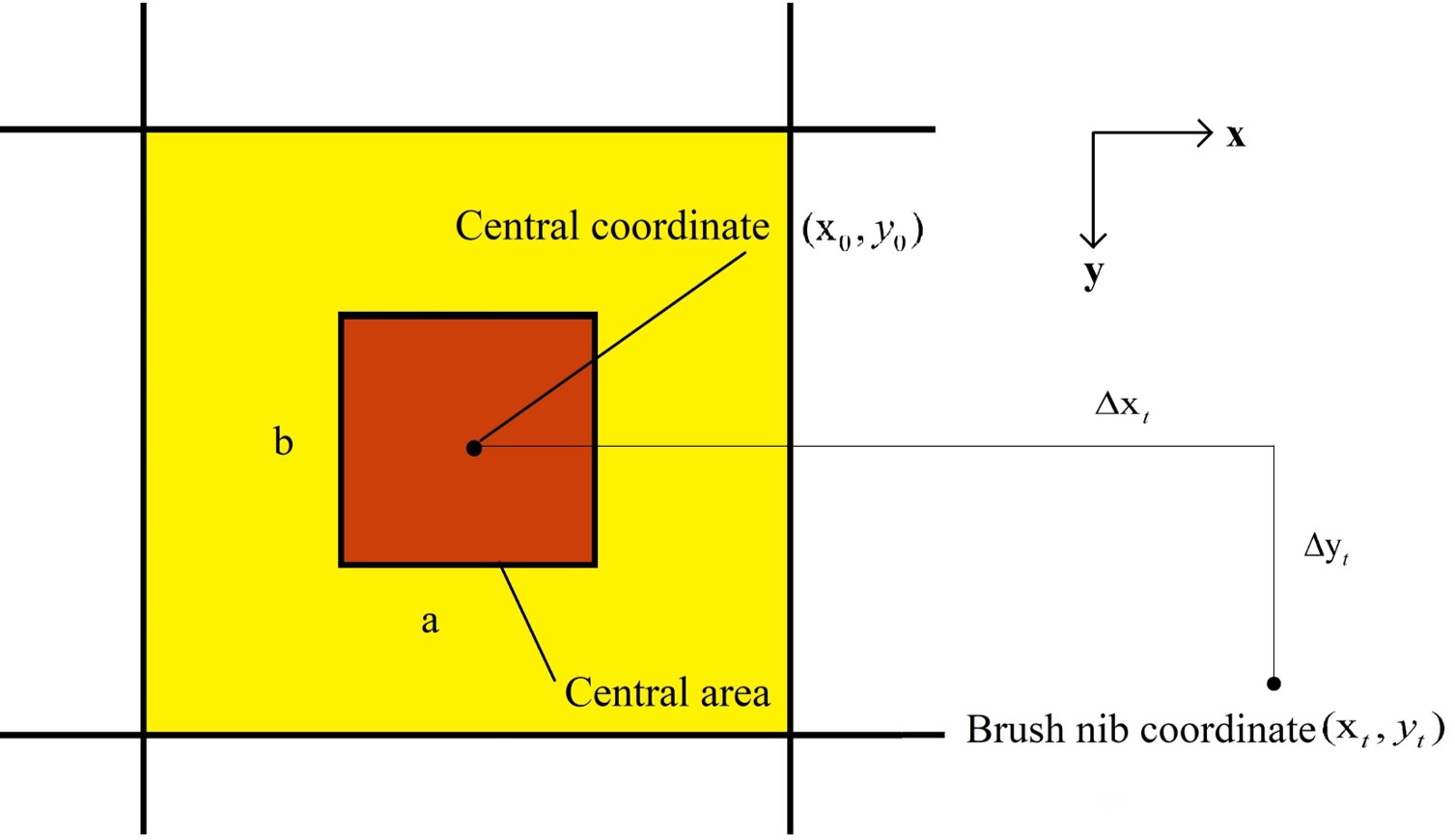}
  \caption{A micro-scenario of the brush tip's arriving at target monitoring. The legend indicates the positive direction. The yellow area denotes the target area while the red area represents central area. ${\Delta x}_{t}$ represents ($x_{t}-x_{0}$) and ${\Delta y}_{t}$ equals ($y_{t}-y_{0}$). If ${2\Delta x}_{t}\leq\,a$ and ${2\Delta y}_{t} \leq\,b$, the brush tip reaches the target successfully at the moment $t$. }\label{A13}
\end{figure}

In the process of drawing, there is a movement speed of the brush tip. Therefore, the time delay inevitably exists between the past position of brush tip announced by the painting navigation system via voice and the current position of brush tip detected by the system. This means that by the time the user hears the voice announcement, the pen has already moved out of the targeted area. To fill this time gap as much as possible, we design a new rule for brush tip positioning. Fig. \ref{A13} shows the rule of judging whether the brush tip reaches target area. We choose a central area from the target area as `the reference area'. The length and width of `the reference area' are labeled as $a$ and $b$, respectively. The coordinates of central point and brush tip are set as $(x_{0}, y_{0})$ and  $(x_{t}, y_{t})$, respectively. Assume  ${\Delta x}_{t} =x_{t}-x_{0}$ and ${\Delta y}_{t}=y_{t}-y_{0}$. If ${\Delta x}_{t}$ and ${\Delta y}_{t}$ meet the criterion ${2\Delta x}_{t}\leq\,a$ and ${2\Delta y}_{t} \leq\,b$, the brush tip reaches the target successfully at the moment $t$, or the painting navigation continues. This rule reduces the time delay and is being optimized in our subsequent work.


\subsection{Human-computer Interaction Module}
\begin{figure}
  \centering
  \includegraphics[width=0.5 \textwidth]{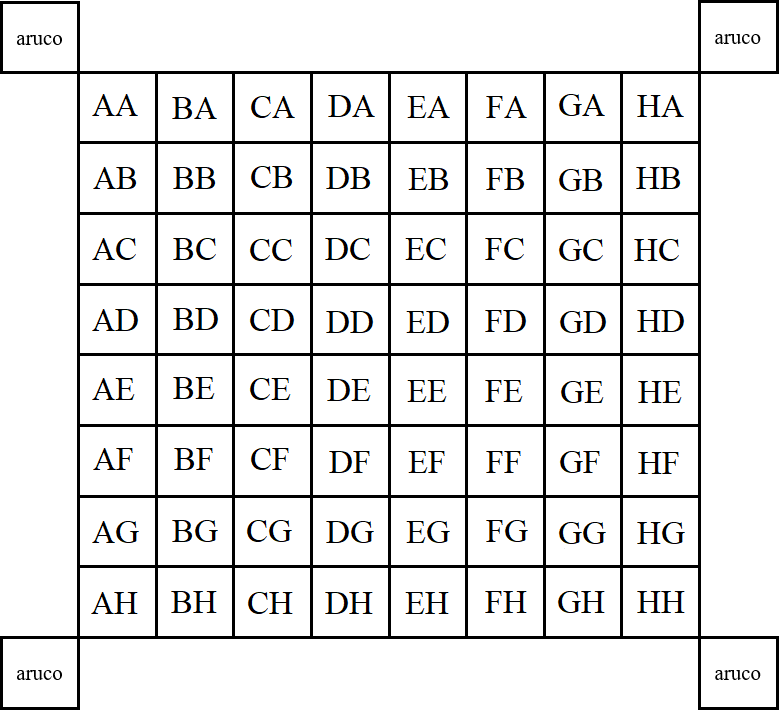}
  \caption{Schematic diagram of panel region segmentation and coding.}\label{A5}
\end{figure}

To allow the blind people to better grasp the location information of the painting area, we use the grid division method to divide the painting area into 64 small independent areas. It should be noted that these sub-regions are programmed by virtual grids, which are invisible for people. The schematic diagram of panel region segmentation and coding is shown in Fig. \ref{A5}. Each small area is coded with two letters. The first and second letters of sub-region represent the row and column where it locates, respectively. For example, the sub-region in the first column and the fourth row is encoded as ``bd", and the sub-region in the fourth column and the fifth row is encoded as ``de".

Based on these virtual sub-regions, the process of human-computer interaction consists of three steps: 1) the user activates the system by the pre-set voice, and presses a button and enters the desired navigation location by voice. The speech recognition unit recognizes the user's voice commands and parses the data to obtain the targeted location information that needs to be navigated; 2) the navigation system gains the coordinate of brush tip using brush positioning module, and then calculates the difference between the coordinates of the target location and the brush tip. Subsequently, four voice prompts namely down, up, left and right are used to guide the user to move the brush to the target location. To avoid confusion, the system only gives guidance information in one direction at a time. For each navigation, the system first navigates in vertical dimension, and then guides the user in horizontal dimension when the vertical and horizontal coordinates of the brush tip are the same. The guidance in one dimension ends, while the guidance in another dimension begins. The above operations alternate back and forth until the targeted region is reached; and 3) when the tip reaches the target position, the system prompts the user to reach the desired painting area by voice.


\subsection{Multimodal evaluation approach}
To verify the practicability of the navigation system, we invite twenty-five blindfolded people, one blind person, and four blindfolded painters who have received professional training in painting to use our painting navigation system. With respect to the testers receiving professional training in painting, we invite them to use the developed system for personalized painting. For the testers without receiving professional training in painting, the target block filling task is designed to evaluate the system performance. The steps of this filling task are shown in Algorithm 1. This procedure is also the workflow of the developed navigation system. The main workflow is: 1) the tester enters coordinate information by voice; 2) the tester looks for the target location by following the voice prompt; 3) the tester fills in with the brush; and 4) the system analyzes the filling condition of the target square to judge whether the color block has been filled. During the experiment, we record the following test information: 1) the thermal video of tester's face; 2) probability distribution of brush tip's occurrence frequency; and 3) the path of brush tip. After the experiment, we conduct the questionnaire survey on the testers. The questionnaire is carefully designed based on empathy map.

The introductions for six evaluation approaches are elaborated as follows:

1) Analysis of facial temperature changes: the thermal camera is used to record the temperature changes of tester in the process of the painting experiment. By analyzing the local temperature changes in the facial area of the testers, we can analyze their emotional fluctuations when using the navigation system. The fluctuation degrees of the temperature of subjects' faces can reflect the user's acceptance of the painting system\cite{tan2002human}.

\begin{figure}
  \centering
  \includegraphics[width=0.45 \textwidth]{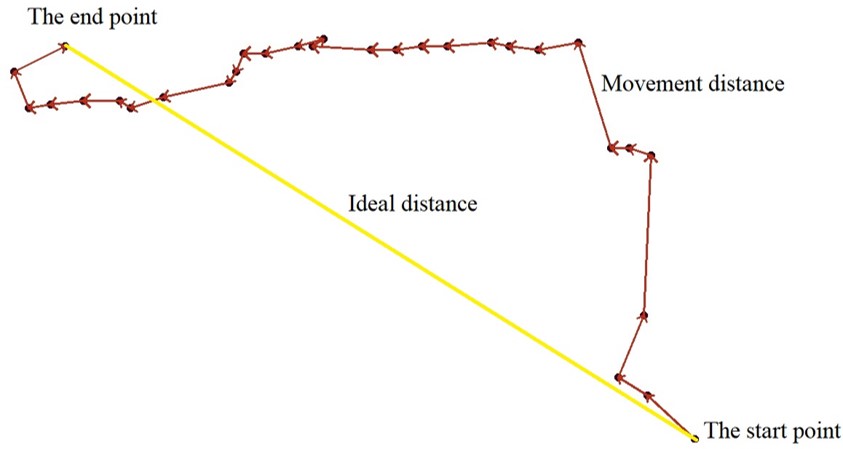}
  \caption{The illustration of `relative movement distance'. The ideal distance is the length of the straightline between the start point and the end point. Additionally, the relative movement distance can be obtained by (\ref{E3}).}\label{A14}
 \end{figure}

\begin{figure*}
  \centering
  \includegraphics[width=1.00 \textwidth]{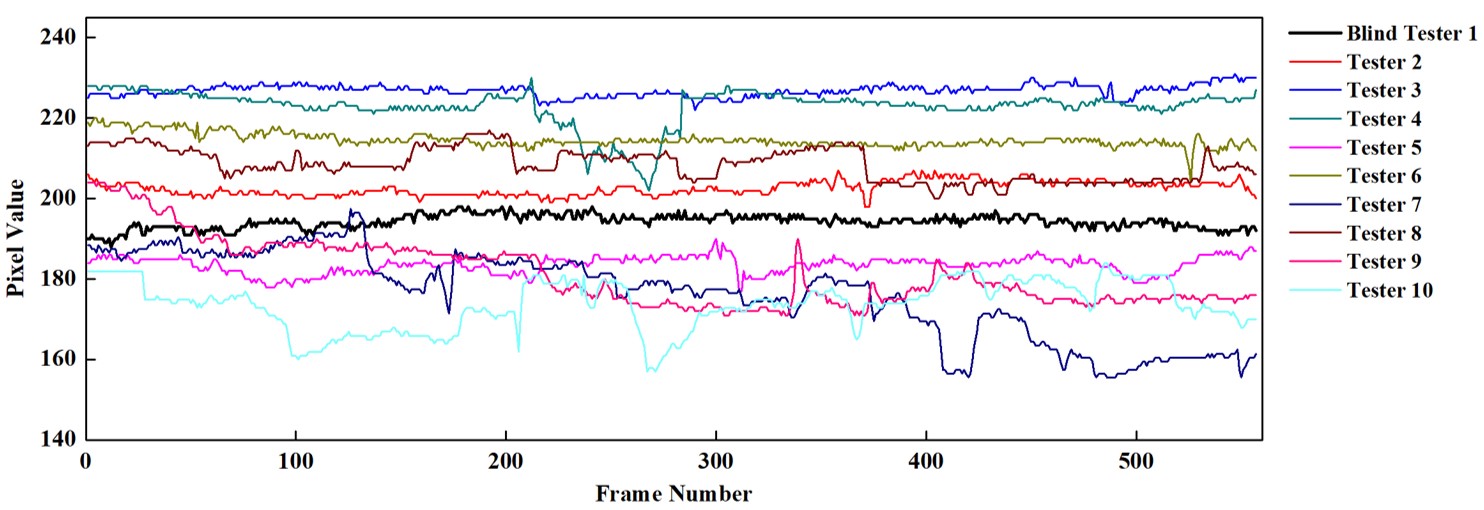}
  \caption{Thermal diagrams of 9 blindfolded testers and 1 blind tester (these thermal curves may reflect their emotional fluctuations when using the navigation system). These thermal curves are derived from 10 testers during two-target destinations experiments using the navigation system. (A total of 13 testers were invited, but in three sets of experiments, the thermal imaging camera malfunctioned or was forgotten to be switched on, so there were three sets of missing thermal imaging data.)}\label{cyz_1}
\end{figure*}

2) Analysis of probability distribution of brush tip: during navigation, the system will record the coordinates of the tip every certain time interval (about 0.3s in this work). After the navigation, the system will automatically generate the probability distribution map of brush tip. This map reflects the occurrence frequency of brush tip: in the painting area, the positions with a high probability are rendered with darker color, while those with a low probability are rendered with lighter color. On the other hand, the brush tip's probability distribution directly indicates the working area involved by the brush during the task: if the task areas are filled with darker color and the other areas are filled with lighter color, the navigation performance is considered to be satisfactory.

3) Analysis of trajectory of brush tip: when the task is completed, the trajectory of brush tip is generated by connecting the positions of brush tip in sequence. Through this trajectory, we can analyze the movement velocity, the movement distance, and movement direction of brush. Intuitively, the distance between two adjacent points is greater, which indicates the movement of brush is faster. Simultaneously, the navigation efficiency can be presented by this trajectory: the shorter the trajectory path is,
the higher efficiency the navigation has.
Furthermore, to make movement distance easier to understand, we introduce `relative movement distance', as elaborated in Fig. \ref{A14}. We set the straight-line distance between the start point and the end point as `ideal distance'. Subsequently, the `relative movement distance' is calculated by:
\begin{equation}\label{E3}
R = \frac{L_{I}}{L_{M}}
\end{equation}
where $L_{I}$ denotes the length of the straightline between the start point and the end point. $L_{M}$ stands for the total length of the brush trajectory. $R$ means the relative movement distance.

Meanwhile, the navigation system is more efficient when the `relative movement distance' is smaller.

4) Analysis of task completion: when the filling area in the target block and the overflowing area account for more than 80$\%$ and less than 375$\%$ respectively, the system considers that the task is completed. The completion degree and the overflow degree are used to evaluate how well the task is done. Particularly, we calculate the overflow degree in the following equation:
\begin{equation}\label{E2}
  O_D = \frac{S_C-S_{T}}{S_T}
\end{equation}
where $S_{C}$ denotes the area of completed block and $S_{T}$ represents the area of target block. $O_D$ means overflow degree.

The completion degree can be calculated using:
\begin{equation}\label{E003}
  O_C = \frac{S_R}{S_T}
\end{equation}
where $O_D$ means completion degree. $S_{R}$ and $S_{T}$ represent the area of completed block in the target area and the area of target block, respectively.

The smaller the value of overflow degree is, the more accurately it is painted. The larger the value of completion degree is, the better the task is completed.

5) Analysis of task completion time: we set three different prompt frequencies (high frequency: 1s; middle frequency: 2s; low frequency: 3s), and record the task completion time of each frequency. The task completion time is long, which indicates the navigation performance is poor; the completion time is short, which indicates the navigation performance is good. In addition, the influence of prompt frequencies on the performance of painting navigation system is analyzed via the tasks with the different prompt frequencies.

6) Analysis of system satisfaction: based on empathy map, we thoughtfully design the questionnaire to the satisfaction of each tester with the system. You can get the the blank questionnaire at: \url{https://doi.org/10.6084/m9.figshare.12611621.v1}. The questionnaire contents include but are not limited to: basic information about the tester; satisfaction; whether the guiding voice conveyed information clearly enough; whether the system provided an accurate painting navigation; how difficult for you to use this system; how much do you think it will help visually impaired painters.

\section{Results of validation experiments}
The developed navigation system is fairly thoroughly validated with the blindfolded people, the blind person, and blindfolded painters using the multimodel evaluation approaches. The original results of evaluation experiments can be found at: \url{https://doi.org/10.6084/m9.figshare.12611621.v1}. The results of validation are elaborated below.
\subsection{Experience of blindfolded and blind users}
Six multimodal evaluation metrics are used to assess the experience of the developed navigation system for both blindfolded and blind users.

1) Analysis of facial temperature changes: the facial temperature changes of the thirteen users recorded by thermal camera are extracted by image processing algorithm, and the obtained thermal diagrams of ten users are shown in Fig. \ref{cyz_1}. To some extent, these thermal curves can reflect the user's emotional fluctuations during the experiment. The fluctuation degrees of all testers are relatively small except the tester 7's, suggesting the navigation system is relatively well accepted by testers. Furthermore, we compare the thermal curves between the blindfolded and blind users. Intuitively, the emotional fluctuation degree of the blind user is much smaller (Fig. \ref{cyz_1}), which indicates that the blind user can better navigate this painting navigation system. Due to the loss of vision, the blind people generally do things more slowly than the people with normal vision. Accumulated over time, the blind people will be more patient than most people with normal vision. As a result, the blind people will also appear to be ``calmer" when using painting navigation system. This phenomenon is quantitatively and objectively revealed in Fig. \ref{cyz_1}.

\begin{figure}
  \centering
  \includegraphics[width=0.45 \textwidth]{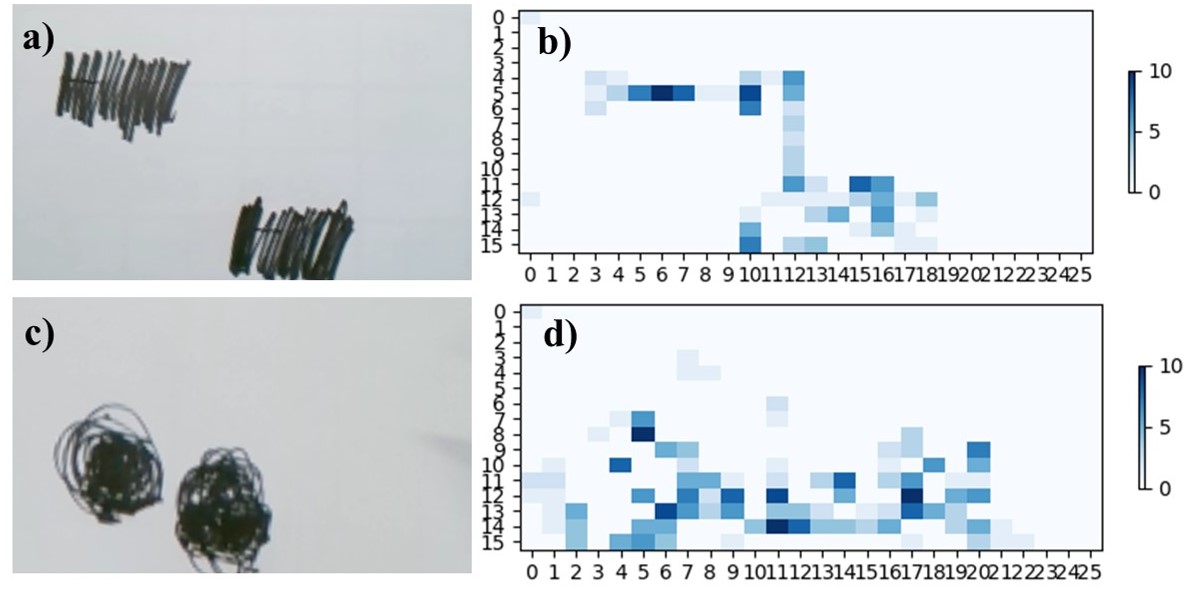}
  \caption{Two typical real drawings and their corresponding brush tip probability distributions for blindfolded and blind users: a) and b) are the real drawing and the corresponding brush tip probability distribution for blindfolded user; and c) and d) are the real drawing and the corresponding brush tip probability distribution for blind user. In the brush tip probability distribution, the length and width of drawing board are divided into 25 blocks and 15 blocks, respectively. Meanwhile, the legend explains the correspondence between color and frequency. The darker the color of block is, the more often the tip appears.}\label{cyz_2}
\end{figure}
2) Analysis of probability distribution of brush tip: the two typical real drawings of blindfolded and blind users and their corresponding brush tip probability distributions are shown in Fig. \ref{cyz_2}. We compare these two sets of data and find it obvious that the probability distribution is in line with the real image. The task areas are rendered with darker color while the others have lighter color, which indicates the performance of the navigation system is satisfactory. Because the traces of the brush tip leave during navigation. It can be observed that a few areas with color are unrelated to the working area. These light color area are reasonable and can be omitted when analyzed. By analyzing all the experimental data, we find that the number of light color regions is related to the distance between the point where the brush enters the drawing board and the target area.

\begin{table}[htbp]
\caption{Average and  standard deviation values of three indicators namely Completion Degree, Overflow Degree, and Relative Movement Distance for two-target experiments.}
\label{table1}
\begin{tabular}{llll}
\toprule
 \multicolumn{1}{m{1.5cm}}{\centering Target} &
 \multicolumn{1}{m{2cm}}{\centering Completion Degree} &
 \multicolumn{1}{m{1.75cm}}{\centering Overflow Degree} &
 \multicolumn{1}{m{2cm}}{\centering Relative Movement Distance} \\
\midrule
 \multicolumn{1}{m{1.5cm}}{\centering Target BC and Target EG} &
 \multicolumn{1}{m{2cm}}{\centering 89$\%$$\pm$8.37$\%$} &
 \multicolumn{1}{m{1.75cm}}{\centering 347$\%$$\pm$162.14\% } &
 \multicolumn{1}{m{2cm}}{\centering 4.21$\pm$2.51} \\

\multicolumn{1}{m{1.5cm}}{\centering Target BC} & \multicolumn{1}{m{2cm}}{\centering 85$\%$$\pm$7.83$\%$} & \multicolumn{1}{m{1.75cm}}{\centering 367$\%$$\pm$198.62$\%$ } & \multicolumn{1}{m{2cm}}{\centering 2.93$\pm$1.44} \\

 \multicolumn{1}{m{1.5cm}}{\centering Target EG} & \multicolumn{1}{m{2.1cm}}{\centering 93$\%$$\pm6.99$$\%$} & \multicolumn{1}{m{1.75cm}}{\centering 325$\%$$\pm$115.41$\%$ } & \multicolumn{1}{m{2cm}}{\centering 6.70$\pm$4.47} \\
\bottomrule
\end{tabular}
\end{table}

\begin{figure*}
  \centering
  \includegraphics[width=1 \textwidth]{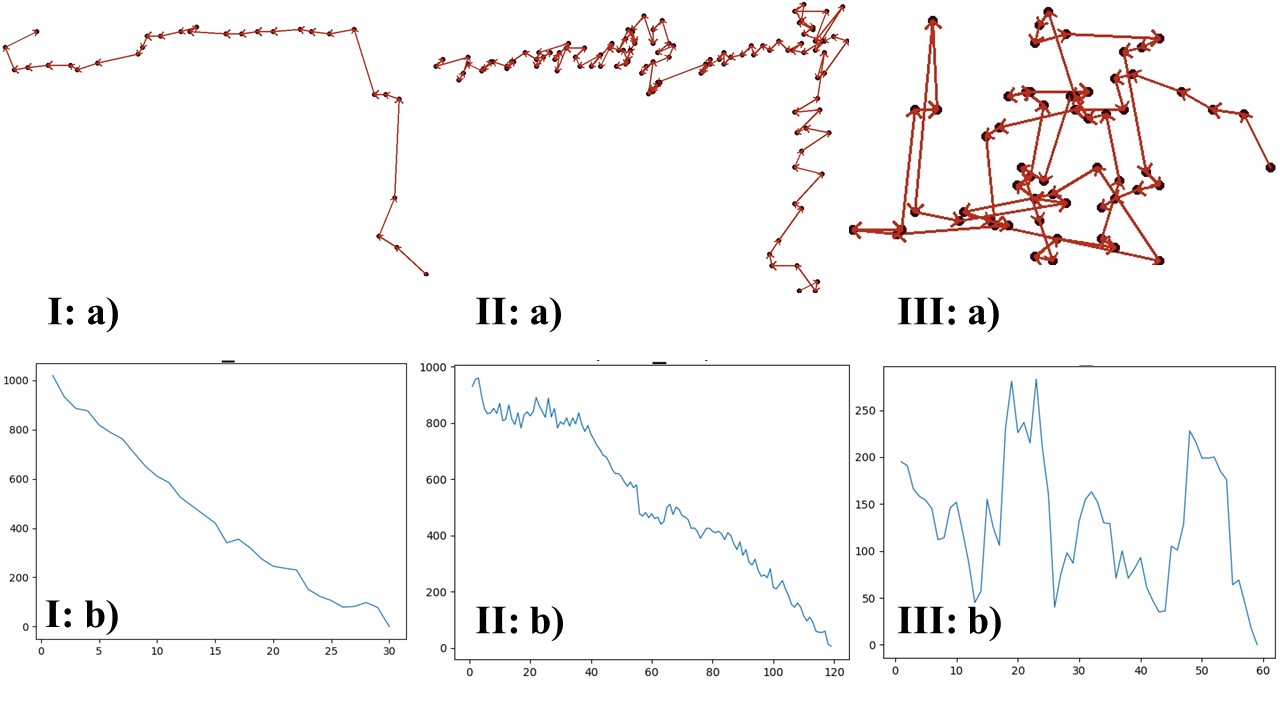}
  \caption{Trajectories of brush tip and their corresponding distance diagrams. Trajectories of brush tip are divided into three categories namely excellent, good and bad according to its trajectory quality: I, II, and III represent the excellent, good and bad categories, respectively; a) and b) represent the trajectory of brush tip and the corresponding distance diagram. The x coordinate represents the serial number of the sampling points (The longer the completion time is, the more the sampling points are.) and the y coordinate denotes the relative distance (The unit of it is pixel.).}\label{cyz_7}
 \end{figure*}
3) Analysis of trajectory of brush tip: all brush tip trajectories are divided into three categories according to its trajectory quality: excellent, good and bad. Subsequently, we obtain the corresponding distance diagram composed of the distance from the sample point in the trajectory to the destination. Fig. \ref{cyz_7} shows the trajectories and distance diagrams of these three typical categories. We can easily get the movement direction of brush and distance change. Obviously, the less messy trajectory is, the more efficient navigation is. In other words, the curve of excellent category in distance diagram is smoother with less fluctuation from the upper left to the lower right. The final results show that the excellent and good types account for 74\% (among 23 valid processed results, the number of type excellent and type good results is 17), indicating the efficiency of the navigation system.

Furthermore, when calculating the relative movement distance, we especially divide the two targets into two parts and then calculate them in three conditions: Target BC and Target EG, Target BC, and Target EG, as shown in TABLE \ref{table1}. It can be noted that the relative distance of target BC is much shorter than target EG' s. To some extent, hoping to finish the experiment as soon as possible, the users move the brush too much, so that the final distance becomes longer. In our experiment, we regard 3.5 as excellent and 4.5 as good. Therefore, the average relative movement distance 4.21 is satisfying. The results will be even better if users experience it with calmer mood. In other words, the navigation system may perform better in realistic operation.

\begin{figure}
  \centering
  \includegraphics[width=0.45 \textwidth]{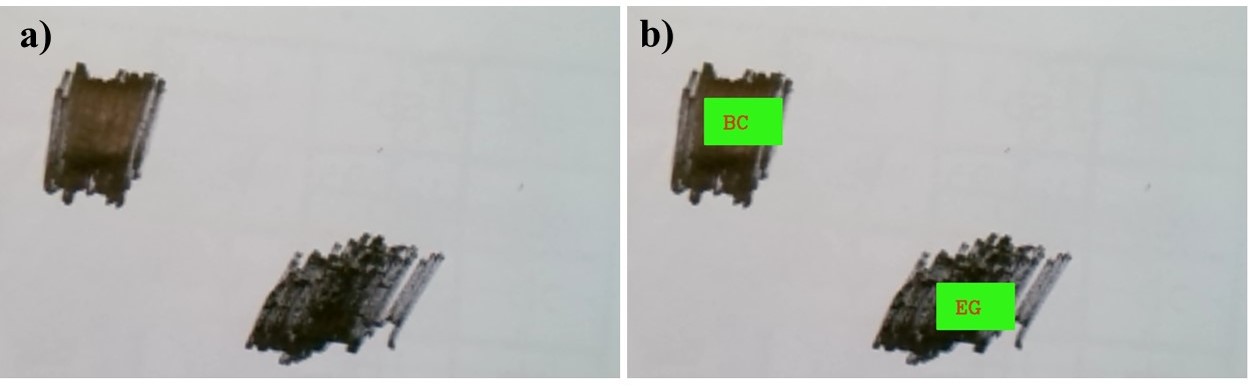}
  \caption{The original drawing and the labeled one. The target area is labeled in every drawing, then the completion ratio and overflow ratio are computed.}\label{cyz_6}
 \end{figure}

4) Analysis of task completion: completion degree and overflow degree of the task are shown in TABLE \ref{table1} while the original and the labeled drawings are shown in Fig. \ref{cyz_6}. Because of the drawing ways of users, almost every target block is overflowed, so we introduce the indicator: overflow degree. The equation is shown in (\ref{E2}).

Notice the completion degrees in all conditions, 89$\%$, 85$\%$ and 93$\%$, are all a lot more than 80$\%$, which suggests the work is well done. The overflow degree meets the completion criterion, 375\%, while the overflow degree of target BC is larger than target EG' s. Considering users have the experience of completing the first target BC, they are more proficient in finishing target EG. Accordingly, the overflow degree is better. The completion degree is excellent and the overflow degree is good while the standard deviation of both indicators is relatively small. Therefore, the navigation system is considered efficient according to the two indicators.

\begin{figure}
  \centering
  \includegraphics[width=0.45 \textwidth]{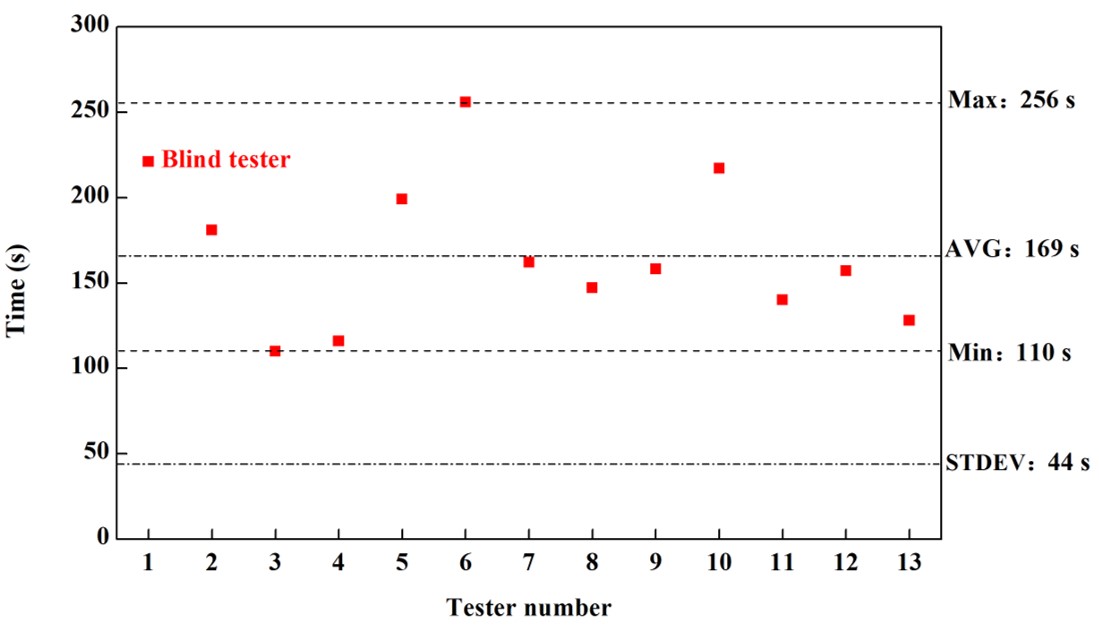}
  \caption{Mathematical statistics of the task completion time: maximal time, minimal time, average time and standard deviation of completion time. Each red point is generated after 13 testers have finished the experiments.}\label{cyz_3}
\end{figure}
5) Analysis of task completion time: in Fig. \ref{cyz_3}, we show completion time of all testers' finishing drawing two target destinations. All results are in the range of 110s to 256s, with the average time 169s and standard deviation 44s, showing the painting navigation system works well for testers. The blind user's time is at a normal level but a little longer than most blindfolded users, which indicates that the painting navigation system may be fit for the blind. One possible reason is that blind people are more patient. Particularly, two testers finished the task within 120s.

6) Analysis of system satisfaction: after sorting questionnaires of the testers, we find the evaluation on satisfaction is positive. It is true of the two following respects namely whether the guiding voice conveyed information clearly enough and whether the system provided an accurate painting navigation. Thus, the results indicate high efficiency of the navigation system. In terms of satisfaction, all testers are satisfied while one third are very pleased. Most users feel easy even very easy to use the navigation system while 13.33\% regard the navigation system a little difficult. Meanwhile, all testers think the system is helpful for blind painters; specifically, the testers thinking it helpful and very helpful account for 28.57\% and 64.29\%, respectively. All feel they can improve proficiency in operating the system through training. In addition, users also raise hope of improvement in the sensitivity of the system perception, the frequency and way of voice guidance, and the robustness of the system, etc.


\subsection{Influence of different prompt frequencies}
\begin{figure}
  \centering
  \includegraphics[width=0.45 \textwidth]{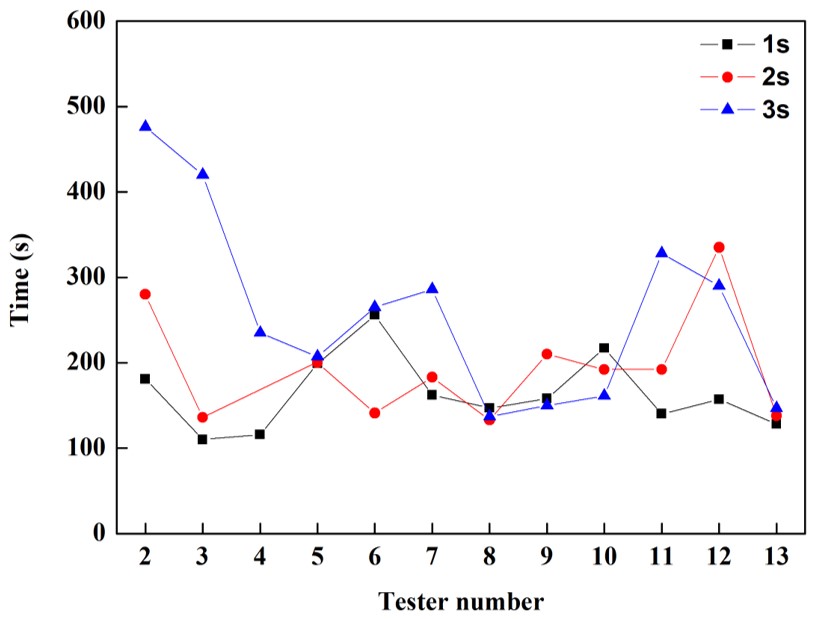}
  \caption{The completion time in three different prompt frequencies. Testers complete drawing two target destinations in prompt frequency of 1s, 2s, and 3s.}\label{cyz_4}
\end{figure}

To validate our system reliability, we carry on experiments with three different prompt frequencies. The results are shown in Fig. \ref{cyz_4}. Among three prompt frequencies, the completion time is the shortest at the high frequency of 1s, followed by the middle, 2s, and the longest at low, indicating the system works best at the frequency of 1s. On the other hand, it can be noted that the task completion time of a few users decreases as prompt frequency increases and the completion time is proximate under three conditions. The ability to accept new things and reaction time varies from person to person, so it is reasonable and acceptable that a few special results appear. Overall, the task completion times for three different prompt frequencies are relatively short.
\subsection{Comparison between different target destinations}
To verify the system adaption, we further invite another 13 blindfolded users to carry out comparison experiments under the condition of three target destinations and the prompt frequency of 1s.

\begin{table}[htbp]
\caption{Comparison of task completion time in different target destination experiments.}
\label{table2}
\begin{tabular}{lllll}
\toprule
 \multicolumn{1}{m{1.25cm}}{\centering Targets Number} &
 \multicolumn{1}{m{1.25cm}}{\centering Maximum time (s)} &
 \multicolumn{1}{m{1.25cm}}{\centering Minimum time (s)} &
 \multicolumn{1}{m{1.25cm}}{\centering Average time (s)} &
 \multicolumn{1}{m{1.5cm}}{\centering Standard deviation (s)}\\
\midrule
 \multicolumn{1}{m{1.25cm}}{\centering Two} &  \multicolumn{1}{m{1.25cm}}{\centering 256} & \multicolumn{1}{m{1.25cm}}{\centering 110} &
 \multicolumn{1}{m{1.25cm}}{\centering 169} &
 \multicolumn{1}{m{1.5cm}}{\centering 44} \\

 \multicolumn{1}{m{1.25cm}}{\centering Three} & \multicolumn{1}{m{1.25cm}}{\centering 385} & \multicolumn{1}{m{1.25cm}}{\centering 123} & \multicolumn{1}{m{1.25cm}}{\centering 261} &
 \multicolumn{1}{m{1.5cm}}{\centering 85}\\
\bottomrule
\end{tabular}
\end{table}
 All the testers succeed in drawing three targets successfully. We show the statistical results in TABLE \ref{table2}. The extra time of average time is about 90s in three-target-destination experiment. Given the greater difficulty, the performance is reasonable and good, which suggests the painting navigation system works well in multi-destination experiment. Meanwhile, the other items namely the maximum and minimum time and standard deviation, show the high efficiency of the painting navigation system.

\subsection{Experience of four blindfolded professional painters}
 \begin{figure}
 \centering
 \includegraphics[width=0.45 \textwidth]{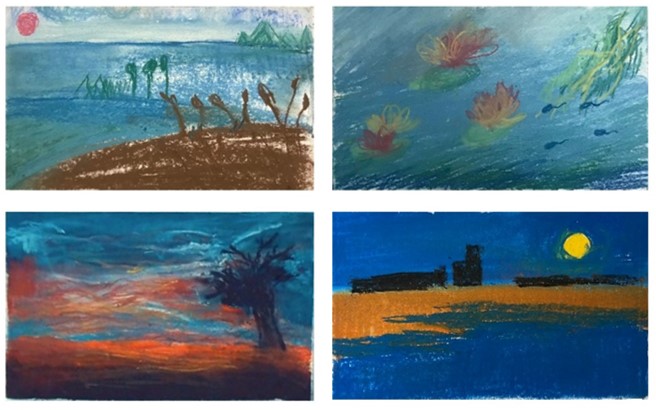}
 \caption{Completed works of professionals with the guidance of the painting navigation system.}\label{cyz_5}
\end{figure}
To validate the actual drawing function of the painting navigation system, we invite four professional painters from the Fine Arts Department to draw under the guidance of the painting navigation system, and the four completed works are shown in Fig. \ref{cyz_5}. The professional painters can finish drawing successfully and create fascinating works, showing the practicability of the painting navigation system. Moreover, the professionals consider the experience of painting satisfactory, which further shows the potential of the navigation system to be actually applied to the blind.
\subsection{Influence of background sound noise on speech recognition accuracy}
To test the noise resistance of the system, we conduct the experiments to analyze the influence of background sound noise on speech recognition accuracy. Speaker devices are used to create different noise conditions. We use a noise detector to measure the noise level at the microphone position. In the range of 19dB $\sim$ 95.3dB, 10 noise levels are selected for this test. ``CD'', ``AB'', ``EF'', ``HG'', and ``FD'' are the target commands in this test. We invite testers to test each instruction four times under each noise condition.

\begin{figure}
 \centering
 \includegraphics[width=0.45 \textwidth]{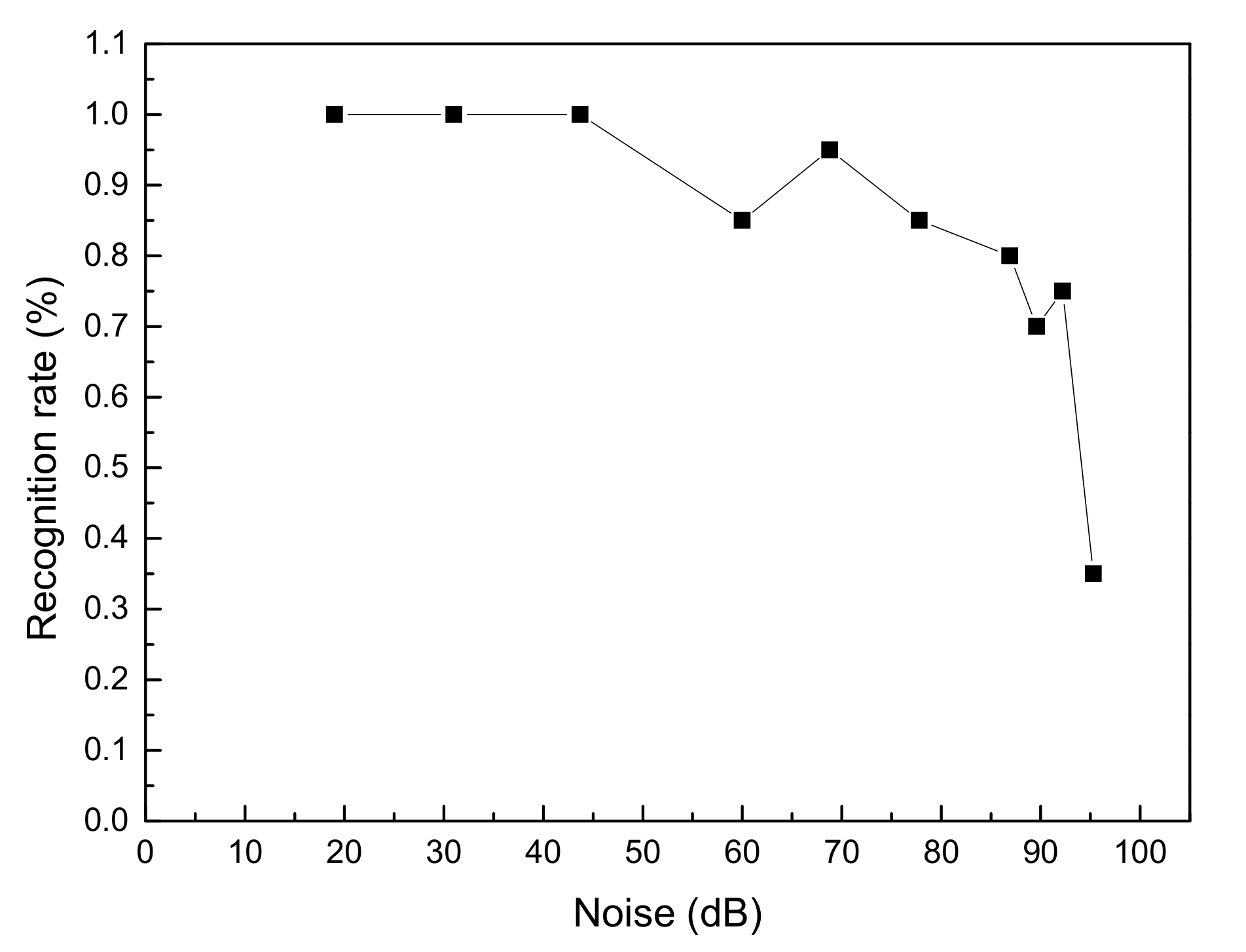}
 \caption{Speech recognition accuracy of the system under different background noise conditions.}\label{A_noise}
\end{figure}

The test results are shown in Fig. \ref{A_noise}. Speech recognition will be affected by environmental noise to a certain extent. When the noise is 43.7dB or less, the recognition accuracy of speech recognition is 100 $\%$. When the noise is above 90dB, the speech recognition accuracy will decrease significantly. The standard of 45 dB indoors in residential areas is identified as the maximum levels below which no effects on public health and welfare occur \cite{noise2021}. According to the test results, the speech recognition accuracy of the system meets the needs when the ambient noise is less than 45dB.

\subsection{Influence of lighting condition on recognition accuracy of brush tip}

We choose five different brightness values between 2LUX and 183LUX in this test. To suppress other factors, the test background is blank. Since the moving speed of brush tip will also affect the test results, the two moving speeds namely the high speed ($>$10cm/s) and the low speed ($<$2cm/s) are set in the test process. The test number of each group is 100.

\begin{figure}
 \centering
 \includegraphics[width=0.45 \textwidth]{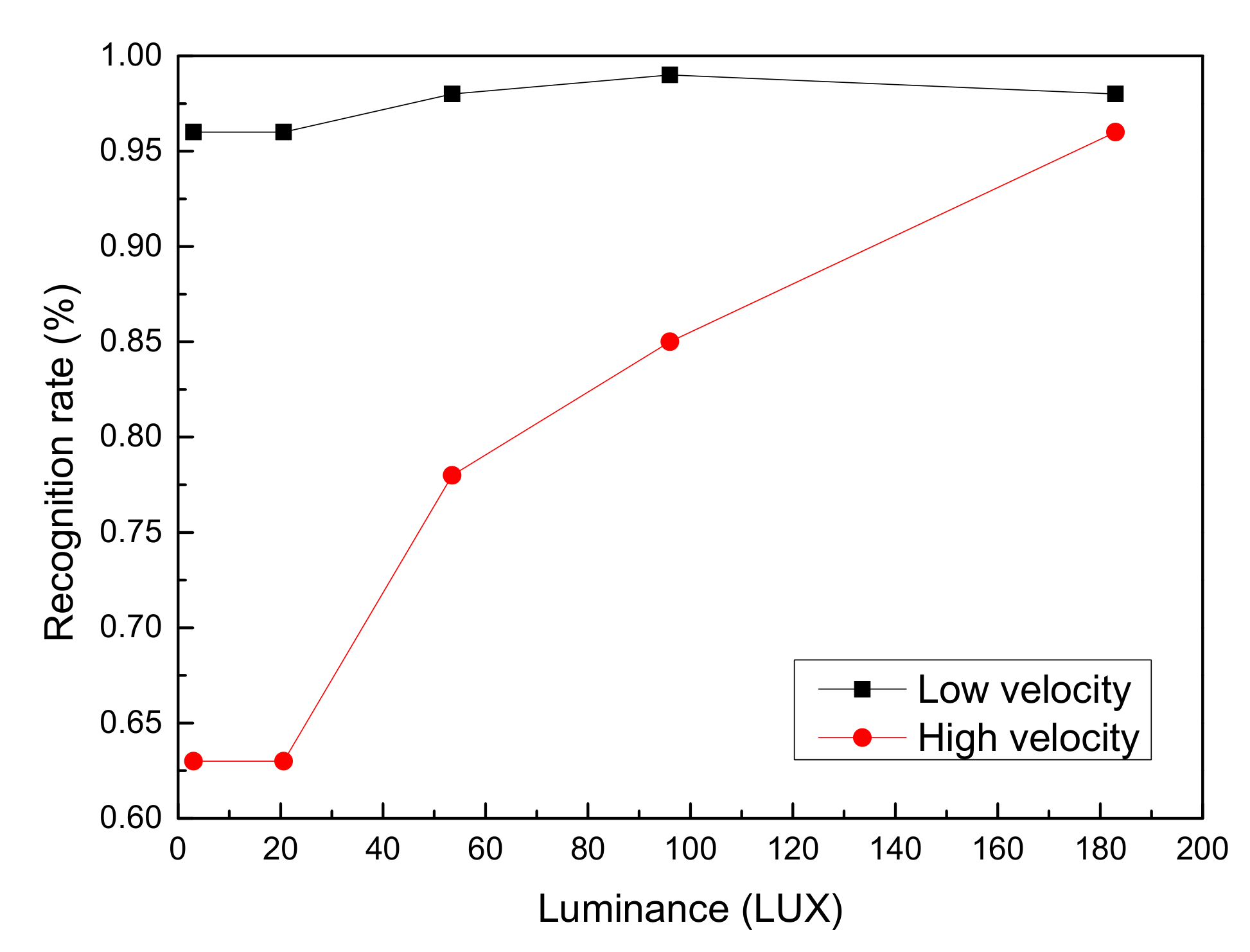}
 \caption{The recognition accuracy of brush tip with the high moving speed or the low moving speed in different lighting conditions.}\label{A_nib}
\end{figure}

    The test results are shown in Fig. \ref{A_nib}. When the brush moves at a low speed, the recognition accuracy of brush tip is above 95$\%$. Under the dark light condition, the recognition accuracy of brush tip for high-speed moving brush tip is between 60$\%$ and 65$\%$. With the increase of brightness, the recognition accuracy of brush tip also increases. When the brightness is 183LUX, the recognition accuracy of brush tip at a high moving speed is 96$\%$. In most applications such as offices and classes, the common light level is around 250LUX \cite{Light2021}. Therefore, the above results indicate that the system can operate under the normal ambient lighting conditions.

\subsection{Influence of complexity of drawing board background on recognition accuracy of brush tip}

We test the influence of the complexity of content in drawing board on recognition accuracy of brush tip. The moving speed of brush tip is set low during the test. Experiments are carried out under five different lighting conditions. We use four paintings (Fig. 15) drawn by blindfolded volunteers using the painting navigation system as the backgrounds in this test. The test number of each group is 100.

\begin{figure}
 \centering
 \includegraphics[width=0.45 \textwidth]{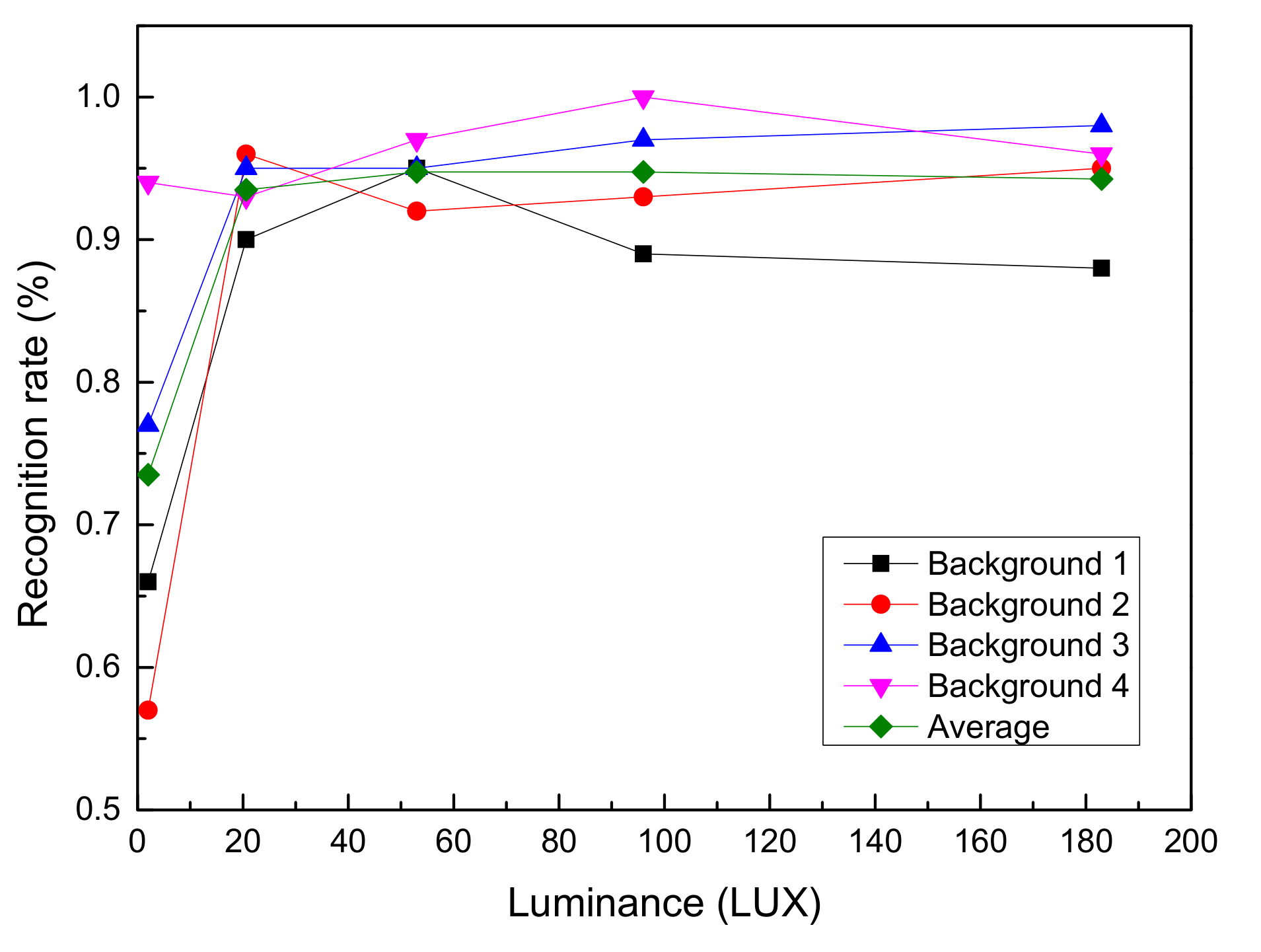}
 \caption{The recognition accuracy of brush tip in the drawing board of complex content and different lighting conditions.}\label{A_background}
\end{figure}

The test results are shown in Fig. \ref{A_background}. In the case of very low illumination, the recognition accuracy of brush tip is low in complex background. When the brightness of the drawing board is normal (\textgreater 20LUX), the average recognition accuracy is more than 90$\%$. According to the above tests, the painting navigation system can work in the drawing board of complex content.

\section{Conclusions and future work}
In this work, we propose an efficient painting navigation system to assist blind painters in painting training and artistic creation. It consists of cognitive system and guidance system. Specifically, the cognitive system is mainly composed of QR code based on drawing board positioning module and brush real-time positioning module. Additionally, we use QR code positioning technology, the dynamic image recognition algorithm, virtual grids to locate the position of the tip, and a digestible two-way voice interaction technology to make the navigation system serve as an “Angel’s Eyes” of blind painters. In mulimodal evaluation, we find the painting navigation system is relatively well accepted by testers especially the blind tester through their thermal curves. With the prompt frequency of 1s, testers can finish the experiment most successfully with the completion degree of 89$\%$ and overflow degree of 347$\%$, showing the efficiency of our system. Analyzing trajectory of brush tip, the type excellent and good account for 74$\%$, and the relative movement distance is 4.12. It is foreseeable that the adaptation will improve after adopting the system and training. Furthermore, the painting navigation system performs well in multi-destination work with the average completion time 261s in three-destination experiment. Most testers believe that they can improve proficiency in operating the navigation system through training.

In the future research, we will aim at improving sensitivity of the system perception, navigation and feedback accuracy. We also plan to deploy the system on the phone to make it more portable.
 \begin{figure}
  \centering
  \includegraphics[width=0.45\textwidth]{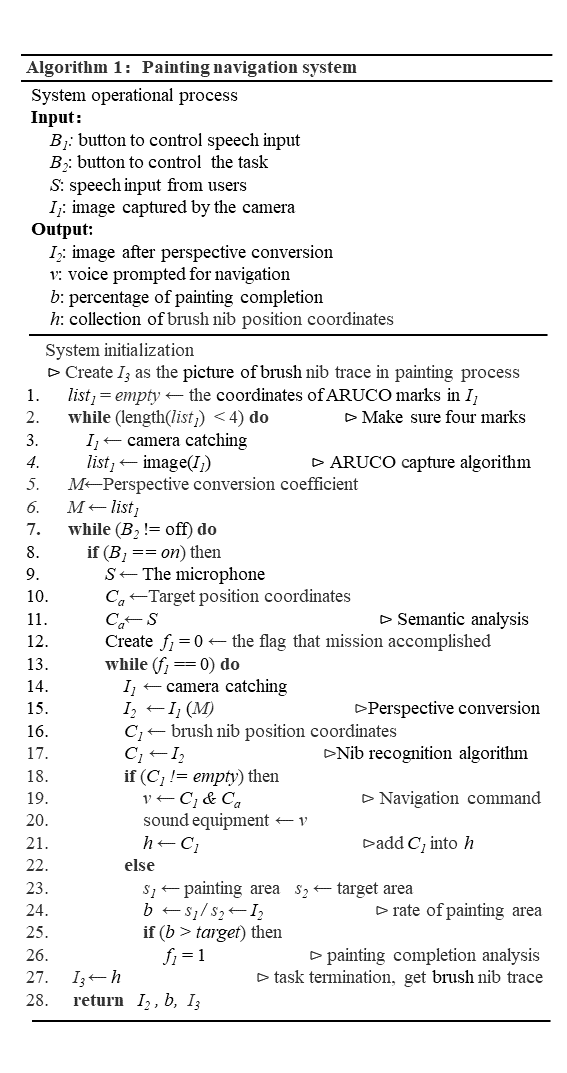}
\end{figure}

%

%
%
%
%
%
%

\ifCLASSOPTIONcaptionsoff
  \newpage
\fi



\bibliographystyle{IEEEtran}
\bibliography{strings,zwh}
\end{document}